latex (revtex) paper and two post script figures follow.

\documentstyle[preprint,revtex]{aps}

\begin{document}


\draft

 \begin{title}
London relation and fluxoid quantization for monopole currents \\
in U(1) lattice gauge theory
 \end{title}

\author{Vandana~Singh,~Richard~W.~Haymaker,~and~Dana~A.~Browne}

\begin{instit}
Department of Physics and Astronomy, Louisiana State University\\
Baton Rouge LA 70803-4001
\end{instit}

\receipt{}

\begin{abstract}
We explore the analogy between quark confinement and the Meissner
effect in superconductors.  We measure the response of color-magnetic
``supercurrents'' from Dirac magnetic monopoles to the presence of a static
quark-antiquark pair in four-dimensional U(1) lattice gauge theory.
Our results indicate that in the confined phase these currents screen
the color-electric flux due to the quarks in an electric analogy of
the Meissner effect.  We show that U(1) lattice gauge theory obeys both
a dual London equation and an electric fluxoid quantization condition.
 \end{abstract}


\narrowtext


Since free quarks have never been isolated, there must be a mechanism
for permanently confining them within hadrons.  This mechanism is
expected to arise in a simple way from the theory of strong
interactions.  It was suggested many years ago\cite{Nielsen} that
confinement would occur if the vacuum reacted to a color electric field
in a manner similar to the response of a superconductor to an external
magnetic field.  Because of the Meissner effect, if two magnetic
monopoles of opposite magnetic charge are introduced into the interior
of a superconductor, the Cooper pairs give rise to persistent currents
to generate a counter magnetic field to expel the magnetic flux.  As a
result, the magnetic flux lines from one monopole to the other are
confined to a narrow Abrikosov flux tube which is surrounded by
persistent circulating currents.  The energy of such a configuration is
proportional to the separation of the two monopoles, thus permanently
confining them.  By analogy, therefore, if the vacuum naturally expels
a color electric flux, the field lines from a static quark-antiquark
$(q\bar{q})$ pair would not spread out in a dipole field pattern but
would instead form a narrow flux tube, leading to a quark potential
proportional to their separation and confinement.

For this mechanism to work the vacuum must contain objects that react
to a color electric field in a fashion similar to the reaction of the
Cooper pairs in a superconductor to an ordinary magnetic field.  One
possibility is to mimic\cite{Nielsen} the Ginzburg-Landau theory of
superconductivity by adding to the gauge theory an elementary Higgs
field to act as the superconducting order parameter, but there is no
experimental evidence for any elementary scalar particles in particle
physics.  The dual superconductor mechanism\cite{Mandelstam} is an
alternative that does not require the ad hoc introduction of a Higgs
field but instead uses dynamically generated topological excitations to
provide the screening supercurrents.  For example, U(1) lattice gauge
theory contains Dirac magnetic monopoles in addition to
photons\cite{Polyakov,Banks}.  The dual superconductor hypothesis
postulates that these monopoles provide the circulating color magnetic
currents that constrain the color electric flux lines into narrow flux
tubes.  't Hooft has shown\cite{AbelProj} that objects similar to the
Dirac monopoles in U(1) gauge theory can also be found in non-Abelian
SU(N) models.

U(1) lattice gauge theory in 4 dimensions has both a confined phase at
large charge and a weak coupling deconfined phase corresponding to
continuum electrodynamics with a Coulomb interaction between static
quarks.  Therefore confinement or its absence can be studied using U(1)
lattice gauge theory as a prototype, before tackling the more
complicated non-Abelian theories that actually describe quarks.  Much
evidence for the dual superconductor hypothesis has accumulated from
studies\cite{Polyakov,Banks,DeGrand,Cardy,Wensley,U1studies} of U(1)
lattice gauge theory.  Polyakov\cite{Polyakov} and Banks, Myerson and
Kogut\cite{Banks} showed that U(1) lattice gauge theory in the
presence of a quark-antiquark pair could be approximately transformed
into a model describing magnetic current loops (the monopoles)
interacting with the electric current generated by the $q\bar{q}$
pair.  DeGrand and Toussaint\cite{DeGrand} demonstrated via a numerical
simulation that the vacuum of U(1) lattice gauge theory was populated
by monopole currents, copious in the confined phase and rare in the
deconfined phase.  This behavior has also been seen in non-Abelian
models after gauge fixing\cite{Kronfeld}.  Many studies of non-Abelian
models using Dirac monopoles\cite{Kronfeld,Suzuki,DSMSU3} or other
topological excitations\cite{SmitIvanenko} support the dual
superconductor mechanism, although other studies\cite{NODSM} dissent.

So far, studies of confinement have examined ``bulk'' properties such
as the monopole density\cite{DeGrand,Kronfeld,Teper}, the monopole
susceptibility\cite{Cardy,Kronfeld} and the behavior of the static
quark potential\cite{Wensley,Suzuki}.  However, the case for the dual
superconductor hypothesis is incomplete without an explicit
demonstration that a static $q\bar{q}$ pair actually induces the
appropriate persistent current distribution.  In this Letter we present
the first direct evidence for this behavior.  We further show that there
are exact U(1) lattice gauge theory analogues of two key
relations\cite{London} that lead to the Meissner effect in a
superconductor; the London equation and the fluxoid quantization
condition.


Our simulations are done on a Euclidean spacetime grid of volume
$L^3\times L_t$, where $L$ is the spatial size and $L_t$ the temporal
size of the lattice in units of the lattice spacing $a$.  The U(1)
gauge degrees of freedom are complex numbers of unit magnitude residing
on the links of the lattice and are written $U_{\mu}(\vec{r}) =
\exp[i\theta_{\mu}(\vec{r})]$, where $\vec{r}$ denotes a point on the
lattice and $\mu$ the direction of the link from that point.  The links
form a directed lattice so that $U_{-\mu}(\vec{r}+\mu) =
U_{\mu}^{\dag}(\vec{r})$.  We use a standard Wilson action $S_o$
supplemented with a Wilson loop $W$ to represent a static $q\bar{q}$
pair with charges $\pm 1$.
 \begin{eqnarray}
   S & = & S_o - iW \nonumber \\
     & = & \beta\sum_{\vec{r},\mu>\nu}
  \left[1-\cos\theta_{\mu\nu}(\vec{r})\right]
     - i \sum_{\vec{r}\mu} J_\mu(\vec{r}) \theta_\mu(\vec{r})
 \label{Action} \end{eqnarray}
  Here $\beta=\hbar c/e^2$ is a dimensionless measure of the strength
  of the charge and $\exp[i\theta_{\mu\nu}(\vec{r})] \equiv
U_\mu(\vec{r})U_\nu(\vec{r}+\mu)U_\mu^{\dag}(\vec{r}+\nu)U_\nu^{\dag}(\vec{r})$
is an oriented product of gauge variables around an elementary
plaquette of the lattice.  The current $J_{\mu}(\vec{r})$ is $\pm 1$
along the world line of the $q\bar{q}$ pair and 0 otherwise.  In the
naive continuum limit $(a \to 0)$ $S$ reduces to the action for a pure
photon field in the presence of a current loop.  Physical observables
are given by expectation values
  \begin{equation}
   \langle A \rangle \equiv {1\over Z} {\rm Tr }\left( e^{-S_o+iW}
   A\right) \label{expectA} \end{equation} where Tr denotes an integral
  over all angles $\theta_{\mu}(\vec{r})$ and $Z = {\rm Tr
}\exp(-S_o+iW)$ is the partition function.

The two observables we study here are the electric flux through a
plaquette, which in lattice variables is ${\cal E}_{\mu}(\vec{r}) =
{\rm Im } \exp[i\theta_{\mu4}(\vec{r})]$, and the curl of the monopole
current density.  The latter is found by a prescription devised by
DeGrand and Toussaint\cite{DeGrand}, which employs a lattice version of
Gauss' Law to locate the Dirac string attached to the monopole.  The
net flux into each plaquette is given by $(\theta_{\mu\nu}(\vec{r})
\bmod 2\pi)$.  If the sum of the fluxes into the faces of a 3-volume at
fixed time is nonzero, a monopole is located in the box.  The net flux
into the box at fixed time thus yields the monopole ``charge'' density,
or the time component of the monopole 4-current $\vec{J}_M$, measured in
units of the monopole charge $g_M=2\pi\hbar c/e$. The spatial components
are found in a similar manner.  The monopole currents form closed loops
due to the conservation of magnetic charge.

Our simulations are performed on a $9^3\times10$ lattice using
skew-periodic boundary conditions.  Less extensive work on a
$7^3\times8$ lattice yields similar results except for the expected
increase in statistical fluctuations arising from the smaller lattice
size.  We used a $3\times3$ Wilson loop oriented in the $zt$ plane and
measured the electric flux and the monopole current in the transverse
($xy$) plane midway along the axis connecting the $q\bar{q}$ pair.  A
standard Metropolis algorithm\cite{Metropolis} alternated with
overrelaxation\cite{relax} is used to generate configurations
distributed according to $\exp(-S_o)$.  In the confined phase, we
thermalize for 10000 sweeps and sample the data every 10 sweeps for a
total of 7000 measurements, which are then binned in groups of 5.  In
the deconfined phase only half as many measurements are taken since the
fluctuations are much smaller.  Because of the geometrical symmetry of
the measurements only the $z$-components of $\langle\vec{{\cal
E}}\rangle$ and $\langle\nabla\times \vec{J}_M\rangle$ are nonzero.  If
the Wilson loop is removed, even the $z$-components average to zero, so
the response is clearly induced by the presence of the $q\bar{q}$
pair.


Figure~\ref{FLUXDIST}(a) shows the electric flux distribution for
$\beta=1.1$ where the vacuum is in the deconfined phase.  The broad
flux distribution seen is identical to the dipole field produced by
placing two classical charges at the quark positions, except that the
classical value of the flux on the $q\bar{q}$ axis is a factor of two
smaller.  We measure the total electric flux from one quark to the
other, including not only the flux through the plane between the
charges ($0.8504\pm0.0045$) but also the flux ($0.0951\pm0.0028$) that
flows through the lattice boundary because of the periodic boundary
conditions.  This yields a total flux of $(0.9453\pm0.0053$), close to
the theoretical value $\Phi_e = e/\sqrt{\hbar c} = 1/\sqrt{\beta} =
0.9534$.

Figure~\ref{FLUXDIST}(b) shows the electric flux in the confined phase
($\beta=0.95$).  In this case the flux is confined almost
entirely within one lattice spacing of the axis and essentially no flux
passes the long way around through the lattice boundary.  The
net flux is again equal to $1/\sqrt{\beta}$ within statistical
error.  This behavior is exactly what one would expect from the
superconducting analogy, where the flux has been ``squeezed'' into a
narrow tube.  The data in Fig.~\ref{FLUXDIST} are consistent with
flux profiles found in other U(1) studies\cite{Burger} and studies of
SU(2)\cite{Sommer} and SU(3)\cite{Burger}.

The electric field $\vec{E}$ produced by the monopole currents arises
from the dual version of Ampere's Law
 \begin{equation}
  -c\nabla\times\vec{E} = \vec{J}_M\label{Ampere}
 \end{equation}
  By analogy to the superconductor, in order to see a dual Meissner
effect we should see a dual London relation\cite{London} between the
field and the current of the form
  \begin{equation}
    \vec{E} = {\lambda^2\over c} \nabla\times\vec{J}_M \label{London}
\end{equation}
where $\lambda$ is the ``London penetration depth'' for the electric
field.  These two equations result in electric fields being confined to
a region of size $\lambda$.

We show in Fig.~\ref{EFLUXJSUM}(a) $\langle\vec{\cal E}\rangle$ and in
Fig.~\ref{EFLUXJSUM}(b) $-\langle\nabla\times\vec{J}_M\rangle$ in the
confined phase as a function of the distance from the $q\bar{q}$ axis.
The data show that the spatial variation of the flux and the curl of
the current are very similar, except for the point on the axis which
will be discussed below.  Figure~\ref{EFLUXJSUM}(c) shows the best fit
found for $\langle\vec{\cal E}\rangle -
(\lambda^2/c)\langle\nabla\times\vec{J}_M\rangle$.  We find a value of
$\lambda/a = 0.482\pm0.008$, which is consistent with the range of
penetration of the electric flux in Fig.~\ref{EFLUXJSUM}(a) and the
thickness of the current sheet in Fig.~\ref{EFLUXJSUM}(b).  The dashed
curve in Fig.~\ref{EFLUXJSUM} is the result of using the continuum
Eqs.~(\ref{Ampere}) and (\ref{London}), and it agrees well with the
flux distribution from the lattice simulations.  We also expect that,
as in a superconductor, the transition to the deconfined phase will be
signalled by a divergence of the London penetration depth.  We have
therefore measured $\lambda$ further from the deconfinement transition
at $\beta=0.90$, and find a smaller penetration depth of
$\lambda/a=0.32\pm0.02$.  In the deconfined phase we find an almost
insignificant value of $\langle\nabla\times \vec{J}_M\rangle$ and
fitted values of $\lambda$ were larger than our lattice size.

The anomalous behavior of the point on the $q\bar{q}$ axis can be
understood by recalling that a superconductor penetrated by an
Abrikosov flux tube becomes multiply connected and the London relation
is replaced by the more general quantization relation for the fluxoid.
Since our U(1) vacuum is pierced by an electric flux tube, we expect a
dual version of the fluxoid quantization relation to hold
 \begin{equation} \int \vec{E}\cdot d\vec{S}-{\lambda^2\over c}
 \oint\vec{J}_{M}\cdot d\vec{\ell} = n\Phi_e, \label{fluxoid}
 \end{equation} where $n$ is an integer and $\Phi_e= 1/\sqrt{\beta}$ is
the quantum of electric flux.  In fact, the data in
Fig.~\ref{EFLUXJSUM}(c) represent a lattice version of a delta-function
whose strength ($1.016\pm0.014$) is very close to
$\Phi_e=1/\sqrt{\beta}=1.026$.  Thus, if the surface integral in
Eq.~(\ref{fluxoid}) includes the axis of the $q\bar{q}$ pair, we obtain
$n=1$, while if the axis is excluded from the integral we obtain $n=0$
and Eq.~(\ref{London}) holds.  We examined the response to
doubly-charged quarks to find $n=2$ electric flux quantization but did
not get data of sufficient quality to draw any conclusions.



Equations~(\ref{London}) and (\ref{fluxoid}) show that, except for the
interchange of electric and magnetic quantities under duality, the
confined phase of U(1) lattice gauge theory behaves exactly like a
superconductor in an external magnetic field.  It is perhaps surprising
that a nonlinear, strongly interacting, model such as U(1) lattice gauge
theory could be described by such a simple model as the linear London
equations, but our results indicate that the operators $\langle
\vec{{\cal E}}\rangle$ and $\langle \nabla\times \vec{J}_M\rangle$ are
clear indicators of the presence of confinement of electric flux by
monopole currents as the dual superconductor mechanism supposes.
Although the Meissner effect itself requires only that
Eq.~(\ref{London}) hold, our data also support the more restrictive
fluxoid quantization relation (\ref{fluxoid}).  Thus the analogy
between Cooper pairs in a superconductor and magnetic monopoles in
gauge theory is very strong.  Analogous studies of SU(2) lattice gauge
theory along these lines are currently in progress.  Because the
monopoles appear pointlike in our simulations, lattice gauge theory
looks like an extreme type-II superconductor and it is tempting to
argue that the phase transition in lattice gauge theory is a Bose
condensation of magnetically charged particles, similar to the Bose
condensation of charged local pairs\cite{Ranninger}.


Two of us (R.W.H. and V.S.) would like to thank G. Schierholz for first
drawing our attention to this problem.  R.W.H. thanks J. Wosiek for
many fruitful discussions on this problem.  V.S. thanks A. Kronfeld, T.
Suzuki and R. Wensley for useful conversations.  R.W.H and V.S. are
supported by the DOE under grant DE-FG05-91ER40617 and D.A.B. is
supported in part by the National Science Foundation under Grant No.
NSF-DMR-9020310.

\figure{Surface plot of the electric flux through the $xy$ plane
midway between the $q\bar{q}$ pair when the system is in (a) the
deconfined phase ($\beta=1.1$) and (b) the confined phase
($\beta=0.95$).  The line joining the pair is located at (0,0).
\label{FLUXDIST}}

\figure{
Behavior of (a) the electric flux, (b) the curl of the monopole current
and (c) the fluxoid in the confined phase ($\beta=0.95$) as a function
of the perpendicular distance $R$ from the $q\bar{q}$ axis.  The dashed
line in (a) shows the flux expected using the continuum relations
(\ref{Ampere}) and (\ref{London}).
\label{EFLUXJSUM}}

\end{document}